\documentclass[prb,letterpaper,noeprint,longbibliography,nodoi,footinbib, twocolumn]{revtex4-1}
\usepackage[colorlinks, allcolors=blue]{hyperref}
\usepackage{amsmath}  % needed for \tfrac, \bmatrix, etc.
\usepackage{amsfonts} % needed for bold Greek, Fraktur, and blackboard bold
\usepackage{graphicx} % needed for figures
\usepackage{subcaption}
\usepackage{tabularx}
\usepackage{verbatim}

\begin{document}

% Be sure to use the \title, \author, \affiliation, and \abstract macros
% to format your title page.  Don't use lower-level macros to  manually
% adjust the fonts and centering.

\begin{widetext}
Notice: This manuscript has been authored by UT-Battelle, LLC, under Contract No. DE-AC0500OR22725 with the U.S. Department of Energy. The United States Government retains and the publisher, by accepting the article for publication, acknowledges that the United States Government retains a non-exclusive, paid-up, irrevocable, world-wide license to publish or reproduce the published form of this manuscript, or allow others to do so, for the United States Government purposes. The Department of Energy will provide public access to these results of federally sponsored research in accordance with the DOE Public Access Plan (http://energy.gov/downloads/doe-public-access-plan).
\newpage
\end{widetext}

\title{Curiosity Driven Exploration to Optimize Structure-Property Learning in Microscopy}
% In a long title you can use \\ to force a line break at a certain location.

\author{Aditya Vatsavai\textsuperscript{1,2}, Ganesh Narasimha\textsuperscript{1}, Yongtao Liu\textsuperscript{1}, Jawad Chowdhury\textsuperscript{1}, Jan-Chi Yang\textsuperscript{3}, Hiroshi Funakubo\textsuperscript{4}, Maxim Ziatdinov\textsuperscript{5}, Rama Vasudevan\textsuperscript{1}}
\email{vasudevanrk@ornl.gov}
%\email{vatsavai23a@ncssm.edu} % optional
%\altaffiliation[permanent address: ]{101 Main Street, 
%  Anytown, USA} % optional second address
%% If there were a second author at the same address, we would put another 
%% \author{} statement here.  Don't combine multiple authors in a single
%% \author statement.
\affiliation{\textsuperscript{1}Center for Nanophase Materials Sciences, Oak Ridge National Laboratory, Oak Ridge, TN, USA - 37831\\\textsuperscript{2}Department of Physics, University of North Carolina, Chapel Hill\\\textsuperscript{3}Department of Physics, National Cheng Kung University, Tainan 70101, Taiwan\\\textsuperscript{4}Department of Materials Science and Engineering, Institute of Science Tokyo, Yokohama, 226-8502, Japan\\\textsuperscript{5}Physical Sciences Division, Pacific Northwest National Laboratory, Richland, Washington, USA – 99352}

%\maketitle
%% Please provide a full mailing address here.
%
%\author{David P. Jackson}

%\affiliation{Department of Physics, Dickinson College, Carlisle, PA 17013}

% See the REVTeX documentation for more examples of author and affiliation lists.

\date{\today}

\begin{abstract}

\section{Abstract}
Rapidly determining structure-property correlations in materials is an important challenge in better understanding fundamental mechanisms and greatly assists in materials design. In microscopy, imaging data provides a direct measurement of the local structure, while spectroscopic measurements provide relevant functional property information. Deep kernel active learning approaches have been utilized to rapidly map local structure to functional properties in microscopy experiments, but are computationally expensive for multi-dimensional and correlated output spaces. Here, we present an alternative lightweight curiosity algorithm which actively samples regions with unexplored structure-property relations, utilizing a deep-learning based surrogate model for error prediction. We show that the algorithm outperforms random sampling for predicting properties from structures, and provides a convenient tool for efficient mapping of structure-property relationships in materials science.
\end{abstract}
% AJP requires an abstract for all regular article submissions.
% Abstracts are optional for submissions to the "Notes and Discussions" section.

\maketitle

\section{Introduction}

Determining structure-property relationships is crucial to the development of new materials with desired functional properties, and therefore rapid determination is critical to accelerate material design and optimization. More generally, in the context of autonomous and self-driving laboratories, rapidly determining the relevant relationships between structure and function is critical to optimizing relevant chemical synthetic and processing pathways for molecular and materials optimization and discovery\cite{tomsdl2024}. 

Microscopy, in particular scanning probe and electron microscopy, provides a powerful method to locally image structures with nanoscale or atomic resolution\cite{bian2021scanning}. In addition, the ability to spatially probe spectroscopic properties allows for correlating the local structure with site-specific functional properties. Traditionally, spatially resolved measurements are performed across a grid of points using techniques such as atomic force microscopy force mapping, scanning tunneling spectroscopy, or electron energy loss spectroscopy in a scanning transmission electron microscope. The downside of this method is that (a) only a small number of points can be probed given a limited experimental time budget, and (b) increasing the number of measured spectroscopic points to increase resolution can result in irreversible tip and/or sample damage. 
Machine learning applications in scientific methods\cite{hase2019next}, especially in the past decade, have impacted imaging techniques\cite{moen2019deep, archit2025segment, krull2020artificial, kalinin2021automated}. Adaptive sampling methods based on route optimization\cite{kandel2023demonstration, godaliyadda2016supervised, noack2019kriging} and sparse sampling\cite{checa2023high, gura2021spiral} have been used for efficient image reconstruction. In particular, with regard to learning structure-property relationships, deep kernel active learning (DKL) approaches have been utilized to adaptively sample material properties using input image patches acquired in the imaging mode on the microscope \cite{liu2022experimental}. This was shown to be highly efficient in correlating local ferroelectric domain structures with specific features of ferroelectric hysteresis loops in the pioneering work by Liu et al \cite{liu2022experimental}. That work was subsequently extended to other modalities, including conductive atomic force microscopy, electron microscopy and scanning tunneling microscopy \cite{liu2023exploring, roccapriore2022physics, narasimha2024multiscale}.  However, DKL, and indeed all Bayesian optimization approaches, utilize a scalarizer function to reduce high-dimensional spectroscopic measurements to a single scalar quantity that is used as the target for optimization\cite{hickman2025atlas}. While this approach is a suitable method to optimize for a given target property, the exploratory power is limited because of the loss of spectroscopic features that are not accounted for by the scalarizer function. Although multi-objective optimization is possible, attempting to develop Gaussian based methods for large output spaces (e.g., above 10 dims) where the outputs are correlated is at present computationally intractable. In principle, ensembles of DKL models for uncorrelated outputs are also a feasible solution, although in practice, spectral outputs tend to be correlated and this strategy is therefore not viable.

Here, we present alternate methods relying on surrogate models of error prediction, which we term curiosity-driven exploration, analogous to the usage of the term in reinforcement learning \cite{zheng2021episodic, thiede2022curiosity, pathak2017curiosity, burda2018exploration}. These methods are based on standard deep neural networks with an encoder-decoder structure that have been employed in the past to predict spectra from images (Im2spec) and images from spectra (Spec2im)\cite{kalinin2021toward}. When the goal is to minimize the loss of an Im2spec or Spec2im model, the optimal scalarizer function is difficult, if not impossible to find. As a solution, we instead determine which spectra to measure by training an auxiliary network to predict Im2spec reconstruction error. The curiosity-driven approach involves sampling regions with high values of the predicted error, so as to rapidly reduce the error of these models.

The paper presents two workflows: The first consists of an ensemble of Im2spec models that is used for spectral prediction, combined with an error model that trains on the spectral mismatch error. In the second method, the error model utilizes the latent space embeddings of an autoencoder to correlate with spectral mismatch. These algorithms, inspired by curiosity-driven reinforcement learning, actively sample spectra for which the structure-property relations have not yet been learned. We first demonstrate and optimize the efficacy of our methods on a pre-acquired dataset. Finally, we implement an algorithm on an atomic force microscope (AFM) to actively learn structure-property relationships in a ferroelectric thin film and discuss possible extensions. 

\begin{figure*}[hbt!]
    \centering
    \includegraphics[width=0.8\textwidth]{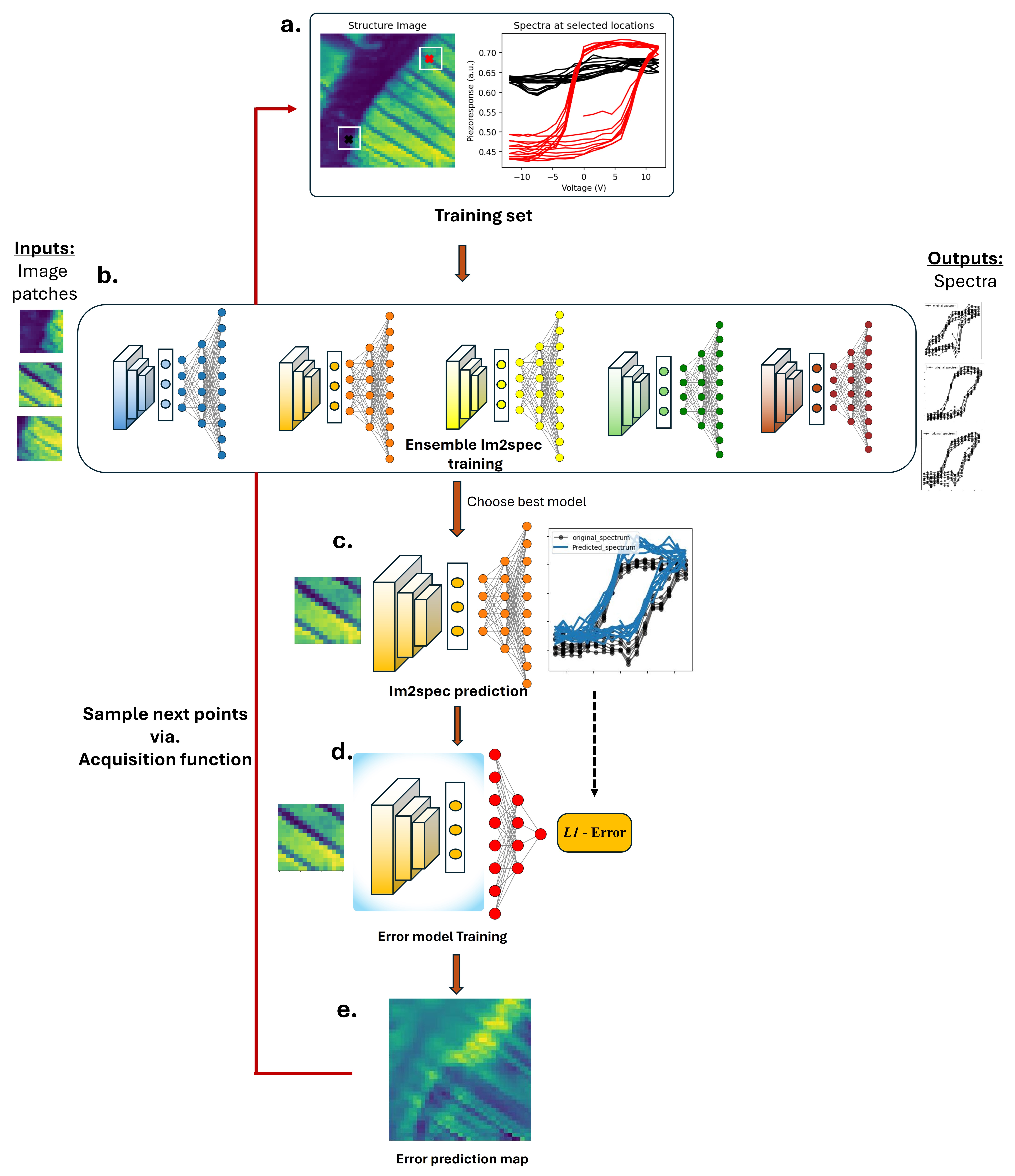}
    \caption{Description of the spectral search method on pre-acquired dataset. \textbf{(a)} Spatial dependence of the spectral property that is correlated with the sample region shown on the left. The patches on the sample image serve as the structural inputs that is correlated to the spectral output. \textbf{(b)} An ensemble of different Im2spec models are trained on an initial set of data. \textbf{(c)} The best model is then used to predict the spectral output corresponding to the training image inputs. \textbf{(d)} Error model consists of the encoder and the latent embedding part of the Im2spec model. This is conjoined with a decoder that is used to train with the spectral mismatch (\textit{L1})error. The parameters of the encoder (and the latent embedding) are frozen while training the error model. \textbf{(e)} The error model is used to estimate errors for the image patches across the sample region. Subsequent sampling points are decided using the acquisition function and incorporated into the training set.}
    \label{fig:enter-label}
\end{figure*}

\section{Results and Discussions}

\subsection{Im2spec encoded error model.}

The structure-property relations described in this work correlate to the ferroelectric response of PbTiO$_3$ samples measured using band excitation piezoresponse microscopy (BE-PFM). We use the structure information from the morphology data acquired from the AFM scanned images, while the property is measured using the spectra collected using band excitation piezoresponse spectroscopy (BEPS) data. This dataset is very similar to one captured and published earlier, and details about the measurement can be found elsewhere\cite{slautin2024bayesian}.

Fig 1 illustrates the active learning workflow described in this section. Fig 1(a) shows sample dataset which shows spatial dependence of the local structure and its influence on the observed spectrum. Here, the local structure, indicated by the square patch, influences the spectrum measured in that region. We initially start by considering a training set where the inputs are the image patches (each patch of size (16 x 16) pixels) while the outputs are the spectra (256 points) corresponding to each patch. In principle, the choice of the patch size is a physics-based quantity which determines the extent of the local structure that affects the measured spectrum. In ferroelelectric measurements, this depends on the electrostatic and elastic fields. We also observe that the window size affects the correlative strength of the input and the output. This can be estimated by comparing the training and the validation loss for the different patch-sizes, results of which are shown in Fig. S1. Low values of the window size results in sub-optimal training while large window size can interfere with efficient learning and result in overfitting.  Our choice of the patch window size (16 pixels) is in the optimal range, with low values of the validation loss. 

In this workflow, we use an ensemble of Im2spec models to offer flexibility for variations in the training data (schematic as shown in Fig 1(b)). Each im2spec model consists of an encoder, a latent embedding layer, followed by a fully connected decoder. While the models are primarily based on the convolutional networks, variations in the architecture have been introduced to enable wider adaptability. A brief description of the encoder architectures used in the model set is provided in Table 1. In our workflow, we designate the size of the latent dimension as three. An initial dataset of the image patches is used to train the Im2spec models. During the training process, we implemented stochastic weighted average for stabilizing the model weights and for generalized spectral prediction. Once trained, the "best model" of the ensemble is chosen based on the minimum validation loss, estimated over last 50 training epochs.

\begin{table}[h]
    \centering
    \begin{tabular}{|>{\centering\arraybackslash}p{0.3\linewidth}|>{\centering\arraybackslash}p{0.6\linewidth}|}
        \hline
        \textbf{Im2spec model name} & \textbf{Encoder architecture} \\
        \hline\hline
        im2spec & Convolution block (3 layers, leaky\_relu = 0.1, dropout = 0.5)\\
        im2spec\_2 & Convolution block (3 layers, leaky\_relu = 0.2, dropout = 0.1) \\
        im2spec\_3& Convolution block (3 layers, leaky\_relu = 0.2,  dropout = 0.1), Dilated block (4 layers) \\
        im2spec\_4 & Resnet module (depth = 3), Convolutional block (3 layers, leaky\_relu = 0.2, dropout = 0.2) \\
        im2spec\_5 & Resnet module (depth = 3), Dilated block (4 layers)\\
        \hline
    \end{tabular}
    \caption{Encoder architecture of the Im2spec models used in the ensemble}
    \label{tab:example}
\end{table}

The selected model is then used to predict the spectral output on the image inputs that were previously used for training (as shown in Fig 1(c)). This prediction is compared with the original spectrum, and the mismatch error is assigned to every image within the training set. We use the \textit{L1} error to quantify the spectral mismatch in this method. Fig 1(d) shows the error model where the Im2spec-encoder (which includes the latent embedding layer) is conjoined with a different set of decoder layers. During the error model training, the encoder part of the model is frozen while the decoder weights are updated. The next step involves the error prediction for the entire set of image patches across the sample region, as shown in Fig 1(e). The error predictions are used to compute the acquisition function to determine and sample the next set of spectral points in an iterative active learning fashion.

Our studies show that the best Im2spec model does not change frequently with minor changes in the training data set. Our code enables probabilistic triggering of ensemble training at selected iterations, helping to avoid redundant training steps. In the results described in this section, we perform ensemble training randomly over 10 \% of the iterations (and the starting iteration). The remaining iterations involve model training using the pre-determined best-Im2spec-model.

Once we predict the errors for all set of the input patches, we use an acquisition function to sample the next data point. The acquisition function used in this method is an empirical equation and is given as: 
$$A_j = 1 - e^{-\lambda|L_j-(1-\beta)|} $$ 
where $L_j$ is the \textit{L1} error normalized in the range [0, 1]. The $\beta$ parameter controls the rate of exploration and exploitation, while the prefactor $\lambda$ controls the smoothness of the acquisition function (higher $\lambda$ indicates better smoothness). The acquisition function varies monotonically with the error values for $\beta$ = 1 and inversely for $\beta$ = 0. This allows us to tune sampling from exploitation to exploration as we increase the $\beta$ hyperparameter from 0 to 1. Therefore the model with $\beta = 1$ is exploration dominant, and is designated as curiosity driven model. In the results described in this section, we study model performance at the extreme, for $\beta$ = 0 and 1 while maintaining $\lambda$ = 0.1.  

The workflow starts with an initial dataset consisting of 245 image-spectrum pairs (20 \%  of the total dataset). Each iteration consists of two model training events - the im2spec ensemble models and the error model. As shown in Fig 1(e), we obtain the prediction of the error values at the end of each iteration. We use the acquisition to sample the next point in every iteration (an alternate method is batch sampling using the acquisition function). In the results described in this section, we study the model behavior over three hundred iterations of active learning.

Fig 2 shows the workflow results where we test the model for the $\beta$ parameter at 0 and 1. We compare the results of the model with a baseline model that trains on acquisitions based on random sampling. Fig 2(a) shows the error statistics for the three models. While the conservative model corresponding to $\beta$ = 0 shows low values of the spectral mismatch error, the curiosity model ($\beta = 1$) shows acquisitions with higher values of the spectral mismatch error. In the curiosity model, at every iteration, new unfamiliar samples improves training over a diverse dataset leading to faster learning. To ascertain this behavior, we estimated the spectral mismatch error over the test set. The results described in Fig 2(b) show a steep reduction of the errors for the curiosity driven model.

Figures Fig 2(c), 2(d), and 2(e) show the acquisition points on the sample region for the random model, $\beta = 0$, and $\beta = 1$, respectively.  We observe that the exploration for the $\beta = 0$ is limited to the domains, and the $\beta = 1$ acquires spectrum in the region of the domain walls and the defective regions of the sample, where the structure to spectral correlations are complex.

\begin{figure*}[hbt!]
    \centering
    \includegraphics[width=0.9\textwidth]{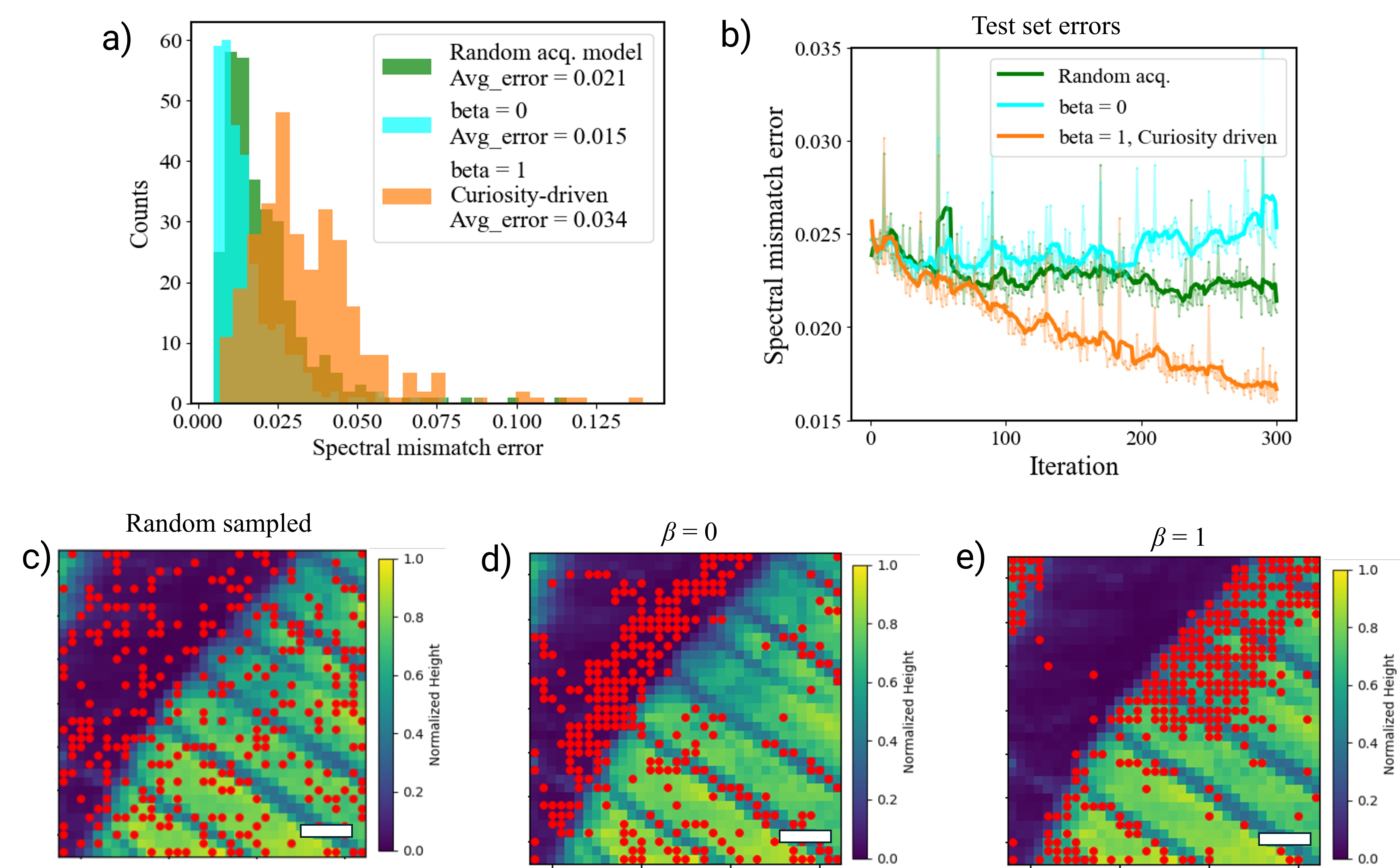}
    \caption{Results of the error prediction model in active learning on the pre-acquired BEPS data. Panel \textbf{(a)} shows the histogram of the error statistics using the acquisition function with random acquisition model,  $\beta$ =0,  and $\beta$ =1. The model with $\beta$ = 1 corresponds to the curiosity driven model that shows sampling of data associated with higher spectral mismatch errors.\textbf{ (b)} Variation of spectral mismatch (\textit{L1}) errors on the test-set for different models. The curiosity based model ($\beta$ = 1) shows steep reduction of the error across the iterations. \textbf{(c)}, \textbf{(d)}, and \textbf{(e)} shows the scatter points where the spectral measurements were acquired across the sample region for the random acquisition model, active learning with $\beta$ = 0, and $\beta$ = 1, respectively .  All scale bars indicate a length of 100 nm.}  
    \label{fig:enter-label}
\end{figure*}

While we see extreme examples of exploitation and exploration for $\beta$ = 0 and 1, respectively, intermediate values of the $\beta$ can be used to balance exploration and exploitation. Fig S2 shows the results of the model and the acquisitions for $\beta = 0.5$.  This samples regions that correspond to both higher and lower values of the error prediction. The explorative performance of the model is therefore intermediate as shown the reduction of the test-set errors in Fig S2(b). 

In the above analysis, and in the rest of the paper, we compare the performance of the model with the commonly used baseline i.e., random sampling. We believe this is reasonable baseline, especially while sampling from a multidimensional dataset. In section 3 of the supplementary material, we show a comparison different sampling techniques and its performance with respect to curiosity based active learning. We observe that the random sampling performs similar to other multidimensional sampling techniques. Nevertheless, curiosity-based active learning outperforms the other sampling techniques.

In an encoder-decoder model, the latent representations that bridge the encoder and the decoder parts of the model determine the efficiency of the reconstruction. We study the latent embeddings to gain insights into the workings of the error model and to interpret the essential features that determine the model output. The latent distributions of the model predictions are described in Fig 3 for the active learning process. Fig 3(a)-(c) represents the latent space distributions for the random model. Fig 3(a) is the latent distribution with the red scatter points, which denote explorations during the active learning process, sampled uniformly. Fig 3(b) shows the latent space clustered into 3 classes, and the corresponding correlation to the real space is shown in Fig 3(c). It is to be noted that the acquisition strategy influences the evolution of the training set, the model weights, and therefore the latent representations. In the random sampling we see uniform sampling across the clusters. Fig 3(d)-(f) shows the latent distribution for  $\beta = 0$. The conservative nature model is reflected in limited exploration that are localized at the high-density region of the latent space. A similar analysis is performed for exploration related to $\beta$ = 1, shown in Fig 3(g)-(i). Here, higher exploration has resulted in a dispersed latent distribution. Further exploration points are comparatively sampled in the sparse region of the latent space. In the real space mapping (Fig 3(i)), this translates to acquisition in the complementary areas (when compared to $\beta = 0$) and corroborates with the data shown in Fig 2(e).

\begin{figure*}
    \centering
    \includegraphics[width=0.9\textwidth]{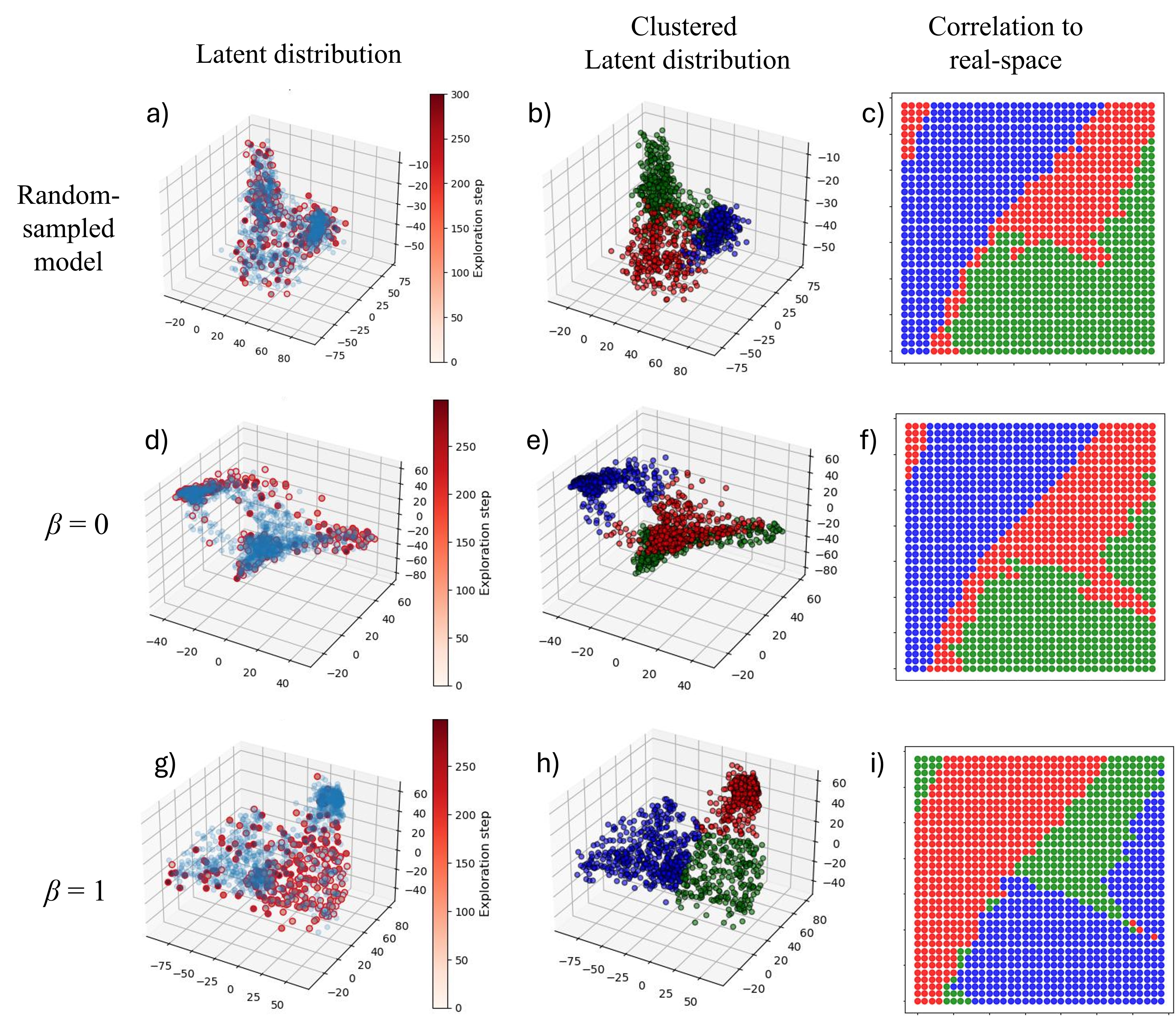}
    \caption{Latent space representations for the active learning-based sampling. Panel \textbf{(a)}-\textbf{(c)} shows the results of the model for randomly sampled model \textbf{(a)} Latent distribution of the model prediction for all image-patch inputs. The red scatter points are the explored samples during the active learning process. \textbf{(b)} depicts the latent distribution clustered into three components using k-means clustering. \textbf{(c)} Correlation to the real-space coordinates corresponding each of the points in the latent space. Panel \textbf{(d)}-\textbf{(f)} shows the similar results for active learning-based sampling with $\beta$ = 0. Figures \textbf{(g)}-\textbf{(i)} show latent distribution results for active learning-based sampling for $\beta$ = 1. While $\beta = 0$ shows predominant sampling in the high dense regions, $\beta = 1$ shows sampling in the dispersed regions of the latent space.}
    \label{fig:enter-label}
\end{figure*}

The results of this section describe the error prediction methods in conjunction with the acquisition function, where the $\beta$ parameter is used to control the degree of exploration/exploitation. At higher $\beta = 1$, the model is curiosity driven and actively seeks unfamiliar samples in the spatial regions of higher predicted error. This allows for diversity within the training set to better learn structure-spectral correlations. 

The latent embedding show uniquely different distributions based on the model and the acquisition strategy. These embeddings serve as compressed, structured representations of input data, capturing essential features of the input images. Given this knowledge, in the subsequent section, we implement a generalized methodology to extract latent representations from an autoencoder while efficiently sampling points from the latent space for active learning based acquisitions.

\subsection{Autoencoder-based error model}

\begin{figure*}
    \centering
    \includegraphics[width=0.8\textwidth]{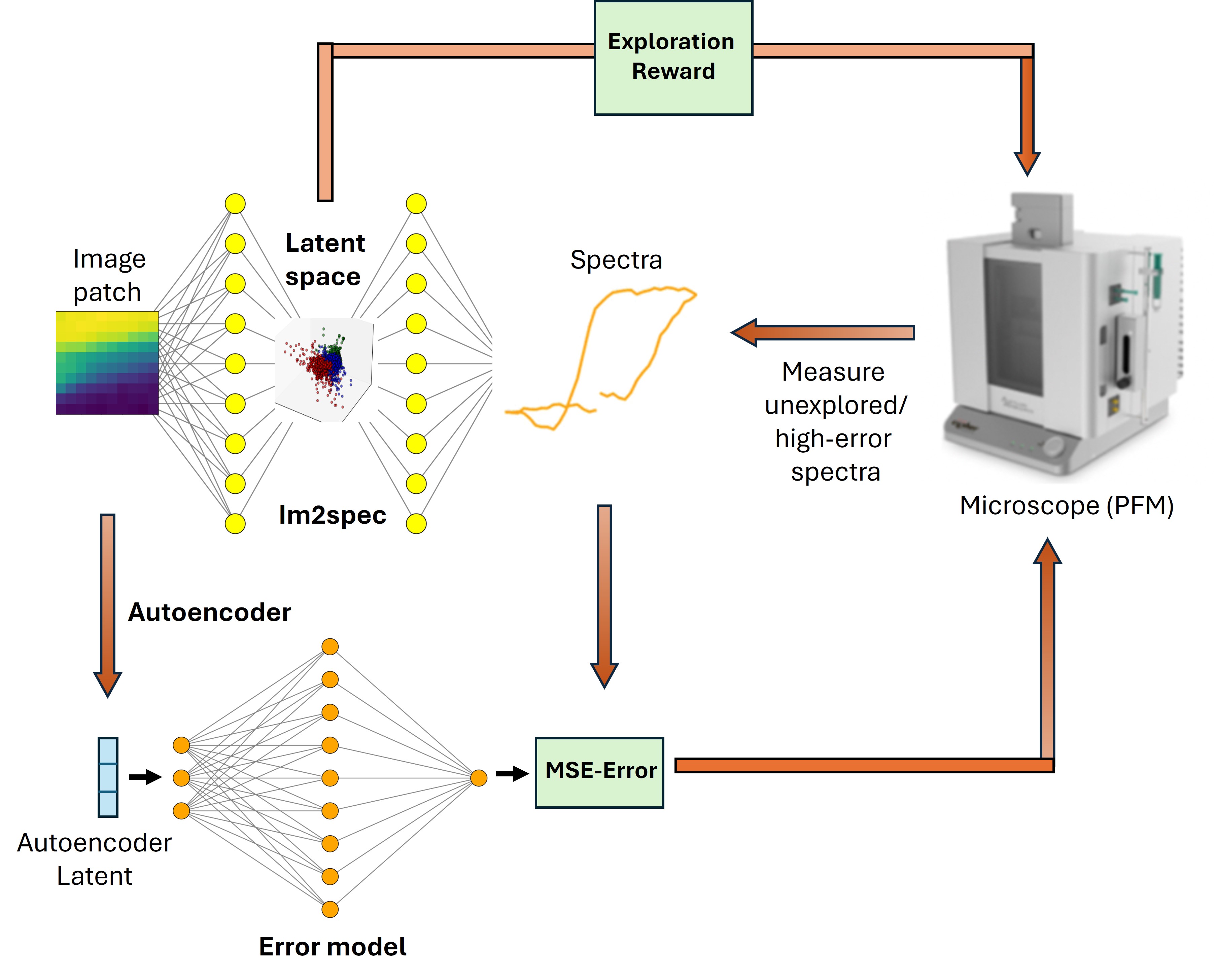}
    \caption{Diagram of Curiosity Algorithm implementation with Im2spec used in conjunction with the Autoencoder-based error model}
    \label{fig:enter-label}
\end{figure*}
This section describes the autoencoder based error model. Here we use a similar experimental dataset - PFM-based experimental data on a 200 nm (110) $PbTiO_3$ thin film sample grown on $SrTiO_3$. The structural information is contained in the image patches (patch size (11$\times$11 pixels)) with switching spectroscopy spectra (spectrum length: 64) captured at low frequency (off-resonance) using an interferometric atomic force microscope from Oxford Instruments (Vero). The baseline curiosity algorithm works as follows: After the sample is imaged an autoencoder is trained on all image patches. Then spectra are acquired on a small number of random initialization points, which are used to train the Im2spec model. The error predictor is then trained on the image patch latent encodings and the Im2spec mean squared error (MSE) for the initial points. Then, a forward pass through the error predictor is performed for all image patch encodings. The spectra of the point with the highest predicted error is then sampled. This continues iteratively where Im2spec and the error predictor are trained on updated training dataset. The overall workflow is illustrated in Fig 4. It should be noted that the error predictor model utilizes dropout to provide an estimate of the uncertainty on the prediction. 

This algorithm is sensitive to the initialization points. If the initial data is not representative of the larger distribution, the algorithm is prone to getting stuck in a local minima.  The error predictor then poorly estimates the Im2spec error for unrepresented data, and therefore fails to sample certain points optimal for reducing Im2spec loss. Therefore for sparse sampling across the distribution, we train an autoencoder on the image patches and then sample the initialization points that are far apart in the autoencoder's latent space. One choice is to utilize $k$-means clustering in the latent space, with $k$ equal to the number initialization points. This was followed by choosing the points closest to each respective cluster centroid as the initialization points. 

To encourage exploration within the latent representations, we reward points that are far away from previously sampled points in the Im2spec latent space. A natural choice for this exploration reward, $E_j$, is the harmonic mean of euclidean distances in the latent space to previously measured points:
$$E_{j} = \left(\sum_{\text{measured } i} \frac{1}{|\ell_i - \ell_j|} \right)^{-1}$$
where $\ell_i$ denote the Im2spec latent encodings of the image patches. Denoting the error predictions as $C_j$, a viable acquisition function, analogous to the epsilon-decreasing strategy for the multi-armed bandit problem \cite{slivkins2024introductionmultiarmedbandits}, is given by:
$$A_j = (1 - e^{-\lambda n}) C_j + e^{-\lambda n} E_j$$ where $n$ is the number of spectra measured so far. Finally, incorporating uncertainty classification in both the error predictor and Im2spec model, and modifying the acquisition function accordingly, would improve exploration. Due to the high dimensional output of Im2spec, we chose to utilize Monte Carlo Dropout\cite{gal2016dropout} for uncertainty estimation. For the error predictor, other methods such as deep kernel learning or a fully Bayesian final layer are also feasible. The exploration reward and model uncertainty classification are not possible for Spec2im, as the spectra required for a forward pass are not available for unmeasured points. In this case, stochasticity can be simply introduced by randomly sampling points with some probability. 

\begin{figure*}[hbt!]
    \centering
    \includegraphics[width=0.96\textwidth]{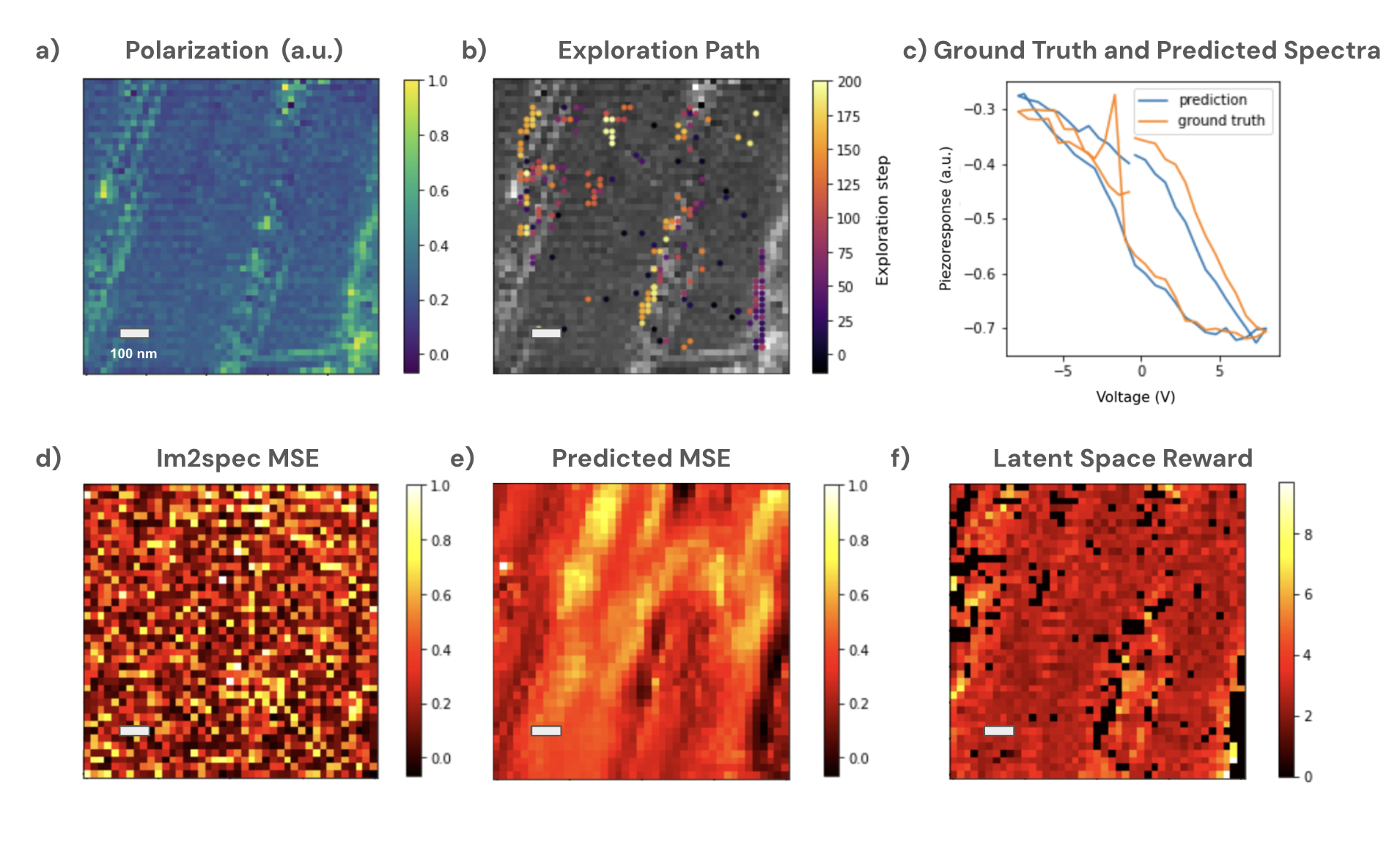}
    \caption{Trial of Im2spec Curiosity Algorithm on pre-acquired PFM data: \textbf{(a)} Polarization ground-truth image, \textbf{(b)} Curiosity algorithm exploration path, and \textbf{(c)} a ground-truth hysteresis loop and corresponding Im2spec prediction. \textbf{(d)} Im2spec MSE error, \textbf{(e)} Predicted error, and \textbf{(f)} Exploration reward after final measurement iteration. Scale bar in the images indicate a length of 100 nm.}
    \label{fig:enter-label}
\end{figure*}
\begin{figure}[b!]
    \centering
    \includegraphics[width=\linewidth]{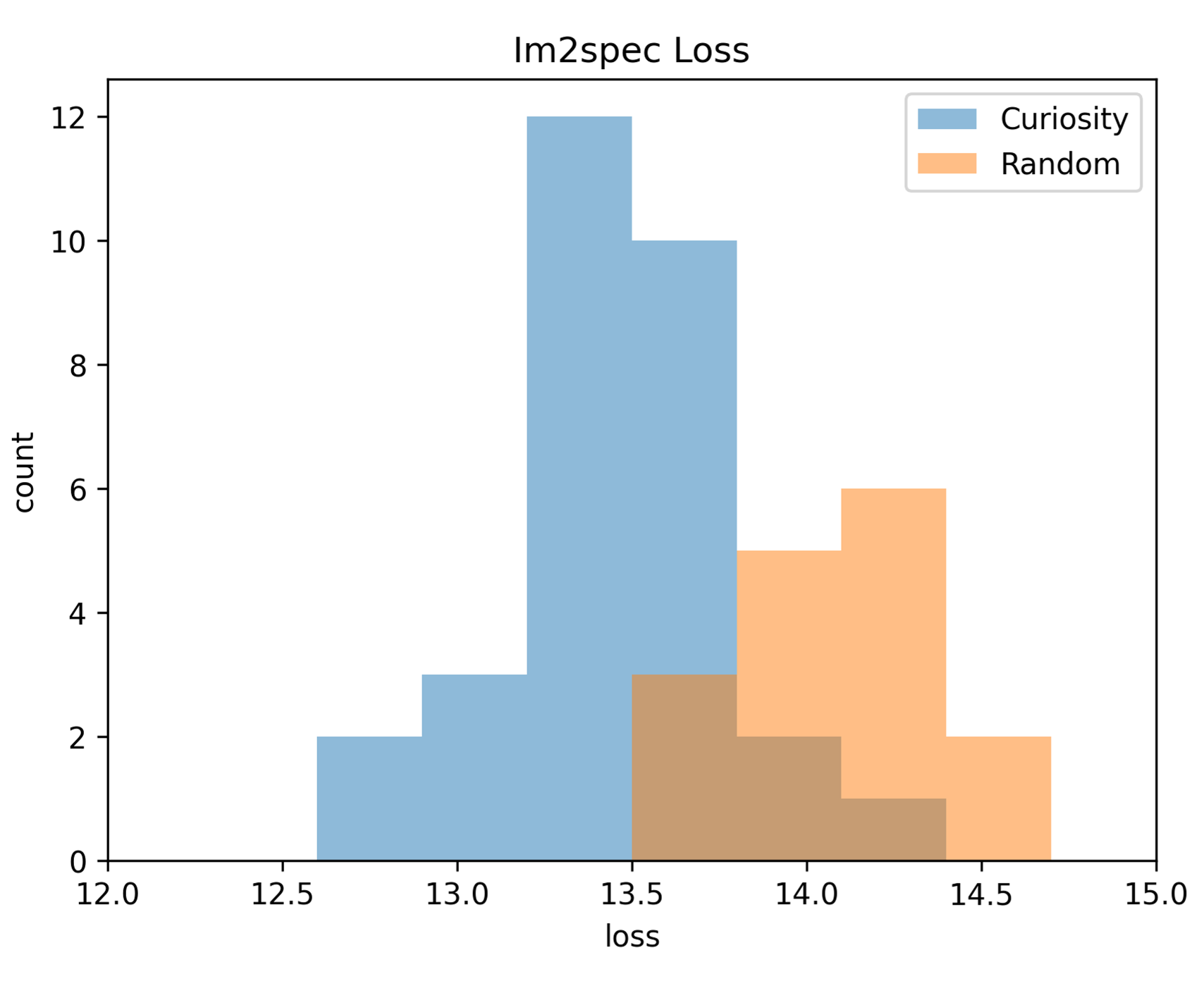}
    \caption{Minimum loss achieved by Im2spec with Curiosity Algorithm vs Random sampling, for 30 trials. The difference between the means is statistically significant}
    \label{fig:enter-label}
\end{figure}
\begin{figure}[b!]
    \centering
    \includegraphics[width=\linewidth]{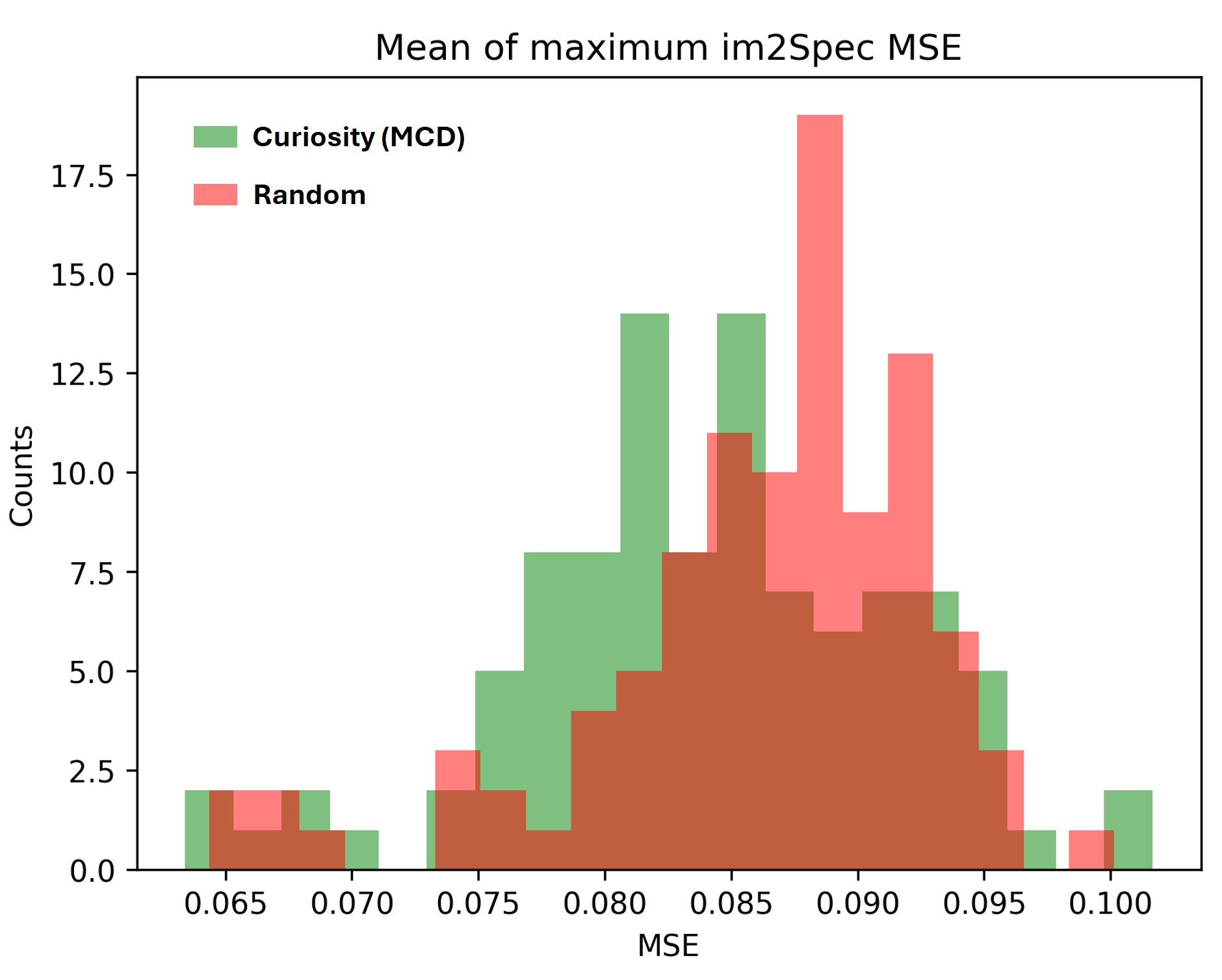}
    \caption{MSE for ten highest error points achieved by Im2spec with MC Dropout Curiosity Algorithm vs Random sampling. The difference between the means is statistically significant.}
    \label{fig:enter-label}
\end{figure}
\begin{figure*}[ht!]
    \centering
    \includegraphics[width=0.98\textwidth]{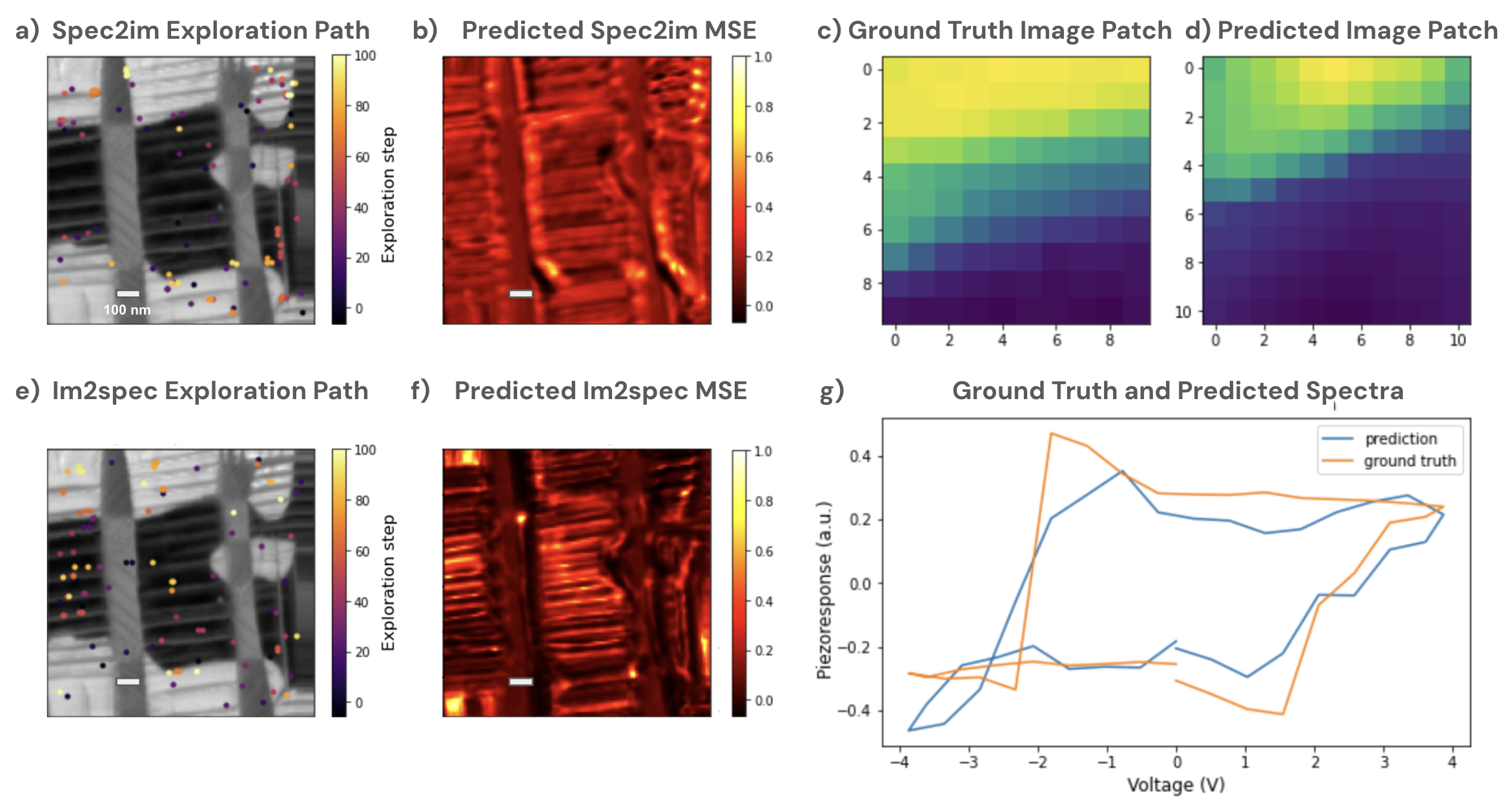}
    \caption{Trial of Curiosity Algorithm real-time on PFM microscope for Spec2im and Im2spec. \textbf{(a)} Exploration path of Spec2im Curiosity Algorithm, \textbf{(b)} predicted Spec2im error after final measurement iteration, \textbf{(c)} a polarization ground-truth image-patch, and \textbf{(d)} corresponding Spec2im prediction. \textbf{(e)} Exploration path of Im2spec Curiosity Algorithm, \textbf{(f)} predicted Im2spec error after final measurement iteration, \textbf{(g)} a ground-truth hysteresis loop and corresponding Im2spec prediction. Scale bar in the images indicate a length of 100 nm.}
    \label{fig:enter-label}
\end{figure*}

Another difficulty is the fact that as Im2spec/Spec2im trains, the MSE values change rapidly. As a result of this non-stationary problem, it is very challenging to train an accurate error predictor. Since the errors decrease on average, the problem can be made more stationary by training the error predictor on the errors divided by the mean error. These normalized MSE errors change much more slowly as Im2spec/Spec2im trains, and allow the error predictor to only account for relative changes in MSE error. It should be noted that even with this modification,  the error predictor required a large learning rate and multiple epochs of training after each measurement in order to keep up with the changing errors.

We tested the Im2spec curiosity algorithm on the aforementioned pre-acquired PFM spectroscopic dataset in order to quantitatively determine its effectiveness. The PFM Polarization image ($P = A\sin(\theta)$), where $A$ is the piezoresponse amplitude and $\theta$ is the phase signal, is shown in Fig. 5(a). We benchmarked curiosity sampling based on predicted error and exploration reward against random sampling. To begin the algorithm, 30 initialization points were seeded, and the algorithm was then run for the next 170 points to sample based on the curiosity metric. The exploration path taken by the algorithm is shown in Fig. 5(b). It is evident that much of the sampling is occurring on the pre-exisiting domain walls, although several clusters of points within the domains are also sampled. The trained im2spec model after the 200 iterations appears to produce decent predictions compared with the ground truth, as shown in Fig. 5(c) for a chosen location. The MSE of im2spec is overall quite low, shown in Fig. 5(d) and does not appear spatially localized. The error predictor predicts maximal errors within the domains, and lowest errors at the domain walls, which also reflects the inverse of the sampled regions, as expected. The exploration reward, after the final measurement iteration, is mapped in Fig. 5(e) and again shows only a few isolated points with high errors. We benchmarked this against random sampling, and the results of the overall loss metrics after running 100 trials are shown in Fig. 6, and show clearly that the curiosity algorithm results in an overall lower loss than random sampling.

In addition, we tested a modified curiosity algorithm which, in addition to latent space exploration reward, samples based on Im2spec Monte-Carlo Dropout (MCD) uncertainty during the exploration phase. While the addition of MCD uncertainty did not directly improve Im2spec loss (the Im2Spec loss was overall not statistically different from a random sampling strategy, in this case), it did reduce the Im2spec MSE for the ten highest error points (Fig. 7). This behavior suggests that enhancing exploration with MCD helps train Im2spec on points with poorly understood structure property relationships, but are not abundantly represented in the sample data, as opposed to points with low error, but are highly represented in the sample data. This, however, has the downside of slower convergence on the whole dataset compared to the original (non-MCD) case. It should be noted that one of the challenges of this algorithm is that there may exist points that continue to contain high errors regardless of the number of training data points, if there are minimal structure-property correlations in these points (for example, if there is only noise in these areas). For such instances, the algorithm should be modified to avoid trapping in these learning plateaus, and strategies can include either direct human intervention, injected noise in the action space, or simple methods such as avoiding similar image patches to past samples if the loss is not decreasing beyond a simple threshold.

\subsection{Real-time microscope deployment}
Given these promising results on pre-acquired data, we moved to implement the curiosity algorithm on the microscope for real-time adaptive sampling. For additional difficulty, we changed the sample to one with a more complex domain structure, a thin film $PbTiO_3$ sample with a hierarchical domain structure that has been previously investigated \cite{liu2022exploring}, and implemented the method using  our AEcroscopy platform for microscope automation and acquisition\cite{liu2024aecroscopy}. Here, we tested both the spec2Im as well as the inverse, Im2spec, for the curiosity algorithm, and plot the results in Fig. 8. The exploration path the algorithm took for the Spec2Im case is shown in Fig. 8(a), and indicates a diverse spread of points across multiple different domain structures. Predicted errors are still spatially localized, but observing examples of predicted images compared with ground truth images show a decent predictive capability (Fig. 8(c,d)). The Im2Spec model shows a different exploration path, with many more points in the darker regions of the image, and the error map appears highly localized, potentially indicating that more points would need to be measured for more accurate modeling. Nevertheless, analysis of the actual predictions shows a decent corroboration with the ground truth (e.g., Fig. 8(g)). 

%We need to add scale bars to the error so we can see whether it is bad or not.

%Results are shown for 30 initialization points and 170 additional measurements for 30 trials. In terms of %Im2spec loss, curiosity sampling outperformed random sampling by a statistically significant amount (). 

%Additionally, we implemented the curiosity algorithm for Im2spec and Spec2im real-time on a PFM microscope. The %exploration paths for Spec2im and Im2spec were substantially different. 

\section{Conclusions}

In summary, we present two different workflows for curiosity driven spectral search. These frameworks utilize latent encodings for error prediction. While the first model utilizes latent space trained for spectral reconstruction, the second autoencoder model describes a generalized approach to train latent embeddings to predict spectral mismatch error. The curiosity algorithm was successful in sampling regions optimal for training Im2spec/Spec2im. On a preacquired dataset, we demonstrated that the curiosity algorithm outperformed random sampling. The algorithm was able to identify regions with complex structure-property relationships, particularly domain boundaries, and preferentially sample these regions in order to minimize Im2spec/Spec2im loss. 

We implemented the workflow on a PFM microscope and found that the exploration paths optimizing Im2spec and Spec2im were different. This discrepancy is fundamentally caused by the in-existence of a bijection between domain structures and hysteresis loops. That is, several structures can produce the same hysteresis loop (for example, structures that are identical apart from a rotation). As a result, a single implementation of the curiosity algorithm is not sufficient for simultaneously optimizing both the forward and inverse problem. In practice, one must choose the algorithm better suited for the given application.

This curiosity based approach is a stepping stone to several novel autonomous microscopy workflows. For example, error prediction can be used to identify regions for which model error is high and does not decrease despite additional measurements, prompting more advanced spectroscopies to be performed in that region. Moreover, the convolutional neural networks may be replaced with theoretical models, in which case the curiosity algorithm would actively sample spectra for which the theory fails, offering insights informing new theoretical models. 

\section{Data Availability Statement }
The data and the code used in this work are available at Zenedo (doi: 10.5281/zenodo.15777800). The Jupyter notebooks can also be found on: https://github.com/cylindrical-penguin/Curiosity-Driven-RL-for-PFM. The scripts presented here utilize open-source repositories \textit{im2spec}(github.com/ziatdinovmax/im2spec) and \textit{atomai} (github.com/pycroscopy/atomai). The deployment of the workflows on the microscopes were interfaced using \textit{aecroscopy} (https://github.com/yongtaoliu/aecroscopy.pyae).

\section{Conflict of Interest}
The authors have no conflict of interest to declare.

\section{Acknowledgment}
Algorithmic development was supported by the US Department of Energy, Office of Science, Office of Basic Energy Sciences, MLExchange Project, award number 107514. The experimental work was supported by the Center for Nanophase Materials Sciences (CNMS), which is a US Department of Energy, Office of Science User Facility at Oak Ridge National Laboratory. GN acknowledges support from QIS Infrastructure Project (FWP ERKCZ62): Precision Atomic Assembly for Quantum Information Science. H.F. was supported by MEXT Program: Data Creation and Utilization Type Material Research and Development Project (No. JPMXP1122683430) and MEXT Initiative to Establish Next-generation Novel Integrated Circuits Centers (X-NICS) (JPJ011438), and the Japan Science and Technology Agency (JST) as part of Adopting Sustainable Partnerships for Innovative Research Ecosystem (ASPIRE), Grant Number JPMJAP2312. J.-C.Y. acknowledges the financial support from the National Science and Technology Council (NSTC), Taiwan, under grant numbers NSTC 112-2112-M-006-020-MY3 and 113-2124-M-006-010.

\bibliographystyle{plain}
\section{References}
\bibliography{references}

\end{document}